\documentclass[aps,superscriptaddress,prd,
onecolumn,
floatfix,
nofootinbib,
longbibliography,
amsmath,amssymb,amsfonts]{revtex4-2}
\usepackage{amsmath,amssymb}
\usepackage{mathrsfs}
\usepackage[dvipdfmx]{graphicx}

\usepackage{bm}
\usepackage{subfigure}
\usepackage{here}
\usepackage{float}
\usepackage{comment}

\newcommand\nn{\nonumber \\}
\newcommand\dd[1]{d{#1}\ }
\newcommand{\pa}{\partial}
\let\Re\relax%Re,Imの定義
\DeclareMathOperator{\Re}{Re}
\let\Im\relax
\DeclareMathOperator{\Im}{Im}
\usepackage[colorlinks,linkcolor=blue,anchorcolor=violet,citecolor=red]{hyperref}

\newcommand{\anno}[1]{\textcolor{red}{#1}}

\begin{document}
%\title{Entanglement Partner in Moving Mirror Models}
\title{The final burst of the moving mirror is unrelated to the partner mode of analog Hawking radiation}
\author{Yuki Osawa}
\email{osawa.yuki.e8@s.mail.nagoya-u.ac.jp}
\affiliation{ Department of Physics, Nagoya University, Nagoya 464-8602, Japan}
\author{Kuan-Nan Lin}
\email{knlinphy@gmail.com}
\affiliation{Leung Center for Cosmology and Particle Astrophysics, National Taiwan University,
Taipei 10617, Taiwan, Republic of China}
\affiliation{Department of Physics and Center for Theoretical Sciences, National Taiwan University, Taipei 10617, Taiwan, Republic of China}

\author{Yasusada Nambu}
\email{nambu@gravity.phys.nagoya-u.ac.jp}
\affiliation{ Department of Physics, Nagoya University, Nagoya 464-8602, Japan}

\author{Masahiro Hotta}
\email{hotta@tuhep.phys.tohoku.ac.jp}
\affiliation{ Department of Physics, Tohoku University, Sendai 980-8578, Japan}
\affiliation{Leung Center for Cosmology and Particle Astrophysics, National Taiwan University,
Taipei 10617, Taiwan, Republic of China}

\author{Pisin Chen}
\email{pisinchen@phys.ntu.edu.tw}
\affiliation{Leung Center for Cosmology and Particle Astrophysics, National Taiwan University,
Taipei 10617, Taiwan, Republic of China}
\affiliation{Department of Physics and Center for Theoretical Sciences, National Taiwan University, Taipei 10617, Taiwan, Republic of China}
\affiliation{Kavli Institute for Particle Astrophysics and Cosmology, SLAC National Accelerator Laboratory, Stanford University, Stanford, California 94305, U.S.A.}

%%%%%%%%%%%%%%%%%%%%%%%%%%%%%%%%%%%%%%%%%%%%%
\date\today
%%%%%%%%%%%%%%%%%%%%%%%%%%%%%%%%%%%%%%%%%%%%%
%%%%%%%%%%%%%%%%%%%%%%%%%%%%%%%%%%%%%%%%%%%%%
\begin{abstract}
 %We consider the entanglement of Hawking radiation in moving mirror models. Hawking radiation (particles) is entangled with their purification partner particles. Based on the partner formula, we carefully investigate the evolution of entanglement in the moving mirror model and discuss that the information recovery based on the vacuum fluctuation scenario does not show the issue of large energy cost, which was pointed out by R. Wald. 
 Flying mirrors with appropriate trajectories have been recognized as an analog system that mimics
black hole Hawking evaporation and have been widely investigated. It has recently been suggested
that the partner mode of the analog Hawking radiation emitted from a moving mirror would
manifest itself through a final burst when the mirror executes a sudden stop. Here we argue
the opposite via the partner formula for the moving mirror model. By expanding the theoretical
foundation of the partner formula and augmenting it with numerical analysis, we demonstrate that
the supposed final burst is induced by a shock that requires the input of external energy, whereas the
Hawking radiation partner mode, which is associated with the zero-point vacuum fluctuations,
is not responsible for the burst.
\end{abstract}
\maketitle
\tableofcontents
%%%%%%%%%%%%%%%%%%%%%%%%%%%%%%%%%%%%%%%%%%%%%

%%%%%%%%%%%%%%%%%%%%%%%%%%%%%%%%%%%%%%%%%%%%%
%%%%%%%%%%%%%%%%%%%%%%%%%%%%%%%%%%%%%%%%%%%%%
%%%%%%%%%%%%%%%%%%%%%%%%%%%%%%%%%%%%%%%%%%%%%
\section{Introduction}
Approximately five decades have passed since Stephen Hawking first proposed the concept of black hole evaporation \cite{Hawking:1975vcx,hawking1974black}. However, the enduring challenge of the information loss paradox \cite{hawking1976breakdown,raju2022lessons} persists in the context of Hawking radiation. According to Hawking's model of black hole evolution, these black holes are formed through the gravitational collapse of massive objects and subsequently undergo evaporation, emitting thermal Hawking radiation. The crux of the issue lies in the fact that even when the initial state is prepared as a pure state before gravitational collapse, the final state evolves to a mixed state, thus violating the principle of unitarity when the black hole experiences complete evaporation. From the perspective of quantum information theory, this violation of unitarity can be understood as the absence of the purification partner of Hawking radiation in spacetime.
One of the most straightforward conjectures to address the information loss paradox is the burst scenario, which posits that the black hole returns the partner of Hawking radiation through some quantum gravitational mechanism once the black hole's mass reaches the Planck scale. However, this scenario implies the emission of vast amount of information at the Planck energy scale, a seemingly implausible proposition.
Numerous researchers have proposed potential candidates for the partner of Hawking radiation that do not entail the energy cost. These candidates include black hole remnants \cite{lin2008entanglement,chen2015black}, Hawking radiation itself as the partner \cite{page1993information,hayden2007black}, vacuum fluctuations \cite{carlitz1987reflections,wilczek1993quantum,holzhey1994geometric,hotta2015partner}, the soft hair of black holes \cite{hotta2001diffeomorphism,hawking2016soft,hotta2018soft}, and, intriguingly, some researchers have explored scenarios that accept information loss at the Planck scale \cite{banks1984difficulties,unruh1995evolution}.

In a recent work by R. Wald \cite{wald2019particle}, the possibility that the vacuum fluctuation scenario is closely associated with the final burst has been highlighted. The basis for his conclusion may be traced to the utilization of the moving mirror model \cite{fulling1976radiation,davies1977radiation}, which occasionally appears in discussions surrounding the vacuum fluctuation scenario.
In moving mirror models, the evaporation of a black hole is represented as the trajectory of a mirror. This trajectory commences with inertial motion, undergoes a constant accelerating phase until it closely approaches a null curve, and subsequently decelerates. During the accelerating phase, the mirror emits thermal radiation, exhibiting similarities to Hawking radiation from black holes. Black hole evaporation is represented by the mirror's deceleration phase, and in this phase a particular issue arises: the radiation flux becomes extraordinarily large when the mirror rapidly decelerates, akin to a burst emission. 
This burst-like radiation is actually a consequence of the external force responsible for the mirror's deceleration. In actual black hole systems, no such external force exists, and the burst phenomenon is not manifest, provided that trans-Planckian physics is not important.
R.Wald considered that the emission of the burst wave can be explained from the perspective of the entanglement monogamy relation. His consideration implies that emission of the burst is an inevitable phenomenon in returning the partner of Hawking radiation during the evaporation process and led him to the conclusion that the vacuum fluctuation scenario is associated with the burst with a large amount of energy.

The main purpose of this paper is to verify Wald's consideration, which is based on the relation between the partner of Hawking radiation and the burst wave.
For this purpose, we have analyzed two types of mirror trajectories. One is the mirror trajectory, which commences with inertial motion, undergoes an accelerating phase until it closely approaches the null curve and then subsequently decelerates to stand still. The other is the mirror trajectory, which initiates in inertial motion, undergoes acceleration, transitions to uniform velocity motion, decelerates, and finally returns to stand still. 
In our examination, we employed the partner formula \cite{hotta2015partner} that describes the spatial profile of the partner.
Then we demonstrate that  the  spatial profile of the partner and its mirror-flipped image at the future null infinity 
%this corresponds to {\color{blue}the partner of the partner} and
({this mirror-flipped image is regarded as the origin of the burst in Wald's paper}) are distinctively separated from the location of the burst.
Our result implies that Wald's description of the burst emission and its relation to partners is not correct, which leads us to conclude that the burst wave is not required to return the partner of Hawking radiation. 
 Therefore, we assert that the vacuum fluctuation scenario differs from the burst scenario.

The organization of this paper is as follows. 
%In Sec.~\ref{sec:revpartner} we briefly review the concept of the partner of the quantum field that appears in \cite{hotta2015partner}. 
In Sec.~\ref{sec:rindlerpartner} we study the Rindler mode and its partner, and visualize them as wave packets in the past null infinity. The wavepackets of Hawking radiation and its partner in the moving mirror model are depicted, and the vacuum fluctuation scenario is reviewed in Sec.~\ref{sec:partner moving mirror}. In Sec.~\ref{sec:wald} we briefly review and comment on Wald's consideration. Sec.~\ref{sec:conclusion} is devoted to the conclusion.

%%%%%%%%%%%%%%%%%%%%%%%%%%%%%%%%%%%%%%%%%%%%%%%%%%
%%%%%%%%%%%%%%%%%%%%%%%%%%%%%%%%%%%%%%%%%%%%%%%%%%%%%%%

%%%%%%%%%%%%%%%%%%%%%%%%%%%%%%%%%%%%%%%%%%%%%%

%%%%%%%%%%%%%%%%%%%%%%%%%%%%%%%%%%%%%%%%%%%%%%%%%%%%%%%%%%%%%%%%%%%%%%%%%%%%%%%%%%%%%%%%%%%%%%%%%%%%%%%%%%%%%%
\section{Rindler Mode and Its Partner in the Flat Spacetime}\label{sec:rindlerpartner}
\subsection{Rindler Mode and Its Partner}
\begin{figure}[H]
	\begin{center}
        \includegraphics[width=7cm]{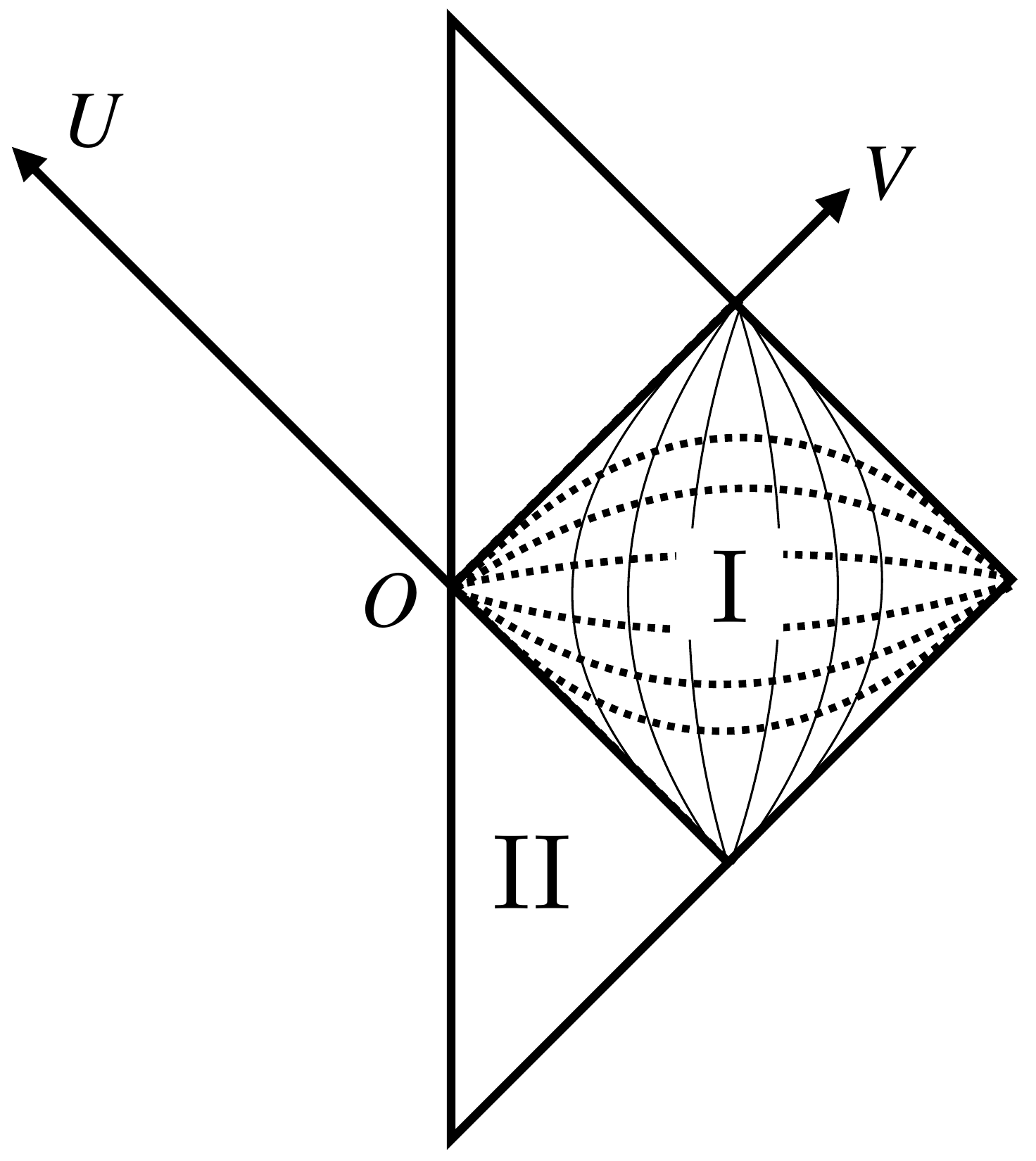}
        \caption{Penrose Diagram of the half of the Minkowski spacetime: solid lines in region I represent $\xi=$ const. lines and dashed lines are $\eta=$const. lines.}
        \label{fig:paraneg}
        \end{center}
\end{figure}
Let us consider a massless scalar field in the $1+1$-dimensional Minkowski spacetime. The scalar field $\phi$ obeys the Klein-Gordon equation $\square\phi=0$. The metric is
\begin{align}
	ds^2=-dt^2+dx^2.
\end{align} 
For later use, we introduce null coordinates, 
\begin{align}
	U=t-x,\ V=t+x.
\end{align}
The metric can be expressed as 
\begin{align}
	ds^2=-dUdV.
\end{align}
As suggested by the equivalence principle, the accelerated observer feels gravitational force, and the spacetime is no longer Minkowskian for him. 
Let $(\eta,\xi)$ be the comoving coordinates for the accelerating observer (Rindler coordinate) where $\eta$ is the time coordinate and $\xi$ is the spatial coordinate; then the Minkowski coordinates and Rindler coordinates are related as
\begin{align}
	t&=\frac{1}{a}e^{a\xi}\sinh a\eta,\quad
	x=\frac{1}{a}e^{a\xi}\cosh a\eta,\quad -\infty<\eta<+\infty,\quad -\infty<\xi<+\infty.
\end{align}
The metric with the Rindler coordinates is given by 
\begin{align}
	ds^2=-e^{2a\xi}(d\eta^2-d\xi^2).
\end{align}
The important property of the Rindler coordinates is that these coordinates cover only region I of the Minkowski spacetime (see Fig. \ref{fig:paraneg}). Thus, the accelerating observer can only see part of the spacetime, and the Minkowski vacuum state is not the vacuum state for the accelerating observer. This property is related to the Unruh effect. We introduce null  coordinates for the Rindler coordinates.
\begin{align}
	u&=\eta-\xi,\quad v=\eta+\xi.
\end{align}
The metric can be expressed as 
\begin{align}
	ds^2=-e^{2a\xi}\,du\,dv.
\end{align}
The null coordinates of the Minkowski spacetime and the Rindler coordinates are related as
\begin{align}
	U&=-\frac{1}{a}\,e^{-au},\quad V=\frac{1}{a}\,e^{av}.
\end{align}
Using the null coordinates, the field equation can be written as $\pa_U\pa_V\phi(U,V)=0$
in the Minkowski coordinates and $\pa_u\pa_v\phi(u,v)=0$ in the Rindler coordinates.
In the Rinder coordinates, the left-moving mode function in region I is defined by 
%%%
\begin{align}
	\phi^{\text{I}}_\omega(V)&:=\exp(-i\omega v(V))\theta(V) \notag \\
	&=\exp\left(-i\frac{\omega}{a}\log(aV)\right)\theta(V),\quad \omega>0,
\end{align}
%%%
 where we multiplied the step function $\theta(x)$
to ensure  the Rindler coordinates  cover only  region I and the mode function $\phi_\omega^\text{I}$ has a support only in region I. This plane wave in the Rindler coordinates is the superposition of plane waves in the Minkowski coordinates with different frequencies:
%%%
\begin{align}
	\phi^\text{I}_\omega(V)=\int_{0}^{\infty}\frac{d\sigma}{{\sqrt{2\pi}}}\left\{\tilde{\phi}^\text{I}_\omega(\sigma) e^{-i\sigma V}+\tilde{\phi}^\text{I}_\omega(-\sigma) e^{i\sigma V}\right\},
	\label{eq:fdecomp1}
\end{align}
%%%
where $\tilde{\phi}^\text{I}_\omega$ is the Fourier component of $\phi_\omega^{\text{I}}(V)$ with respect to $V$.
The Fourier component corresponding to the positive frequency mode can be calculated as 
\begin{align}
	\tilde{\phi}^\text{I}_\omega(\sigma)&=\int_{-\infty}^\infty \frac{dV}{{\sqrt{2\pi}}} \exp\left(-i\frac{\omega}{a}\log(aV)\right)e^{i\sigma V}\theta(V) \notag \\
%	&=\int_{0}^\infty\dd V \exp\left(-i\frac{\omega}
%{a}\log(aV)\right)e^{i\sigma V} \notag \\
	&=i\int_0^\infty\frac{dV}{{\sqrt{2\pi}}} \exp\left(-i\frac{\omega}{a}\log aV\right)e^{\pi\omega/2a}e^{-\sigma V},
\end{align}
%%%
where we have deformed the integral contour on the real axis $V$ into the integral contour on the imaginary axis, and used the relation $i=e^{i\pi/2}$.
Performing the similar calculation we obtain the Fourier component corresponding to the negative frequency mode as
\begin{align}
	\tilde{\phi}^\text{I}_\omega(-\sigma)&=-i\int_0^\infty\frac{dV}{{\sqrt{2\pi}}} \exp\left(-i\frac{\omega}{a}\log aV\right)e^{-\pi\omega/2a}e^{-\sigma V} \notag \\
	&=-e^{-\pi\omega/a}\tilde{\phi}^\text{I}_\omega(\sigma).
\end{align}
Substituting this relation into Eq.~(\ref{eq:fdecomp1}), we obtain
\begin{align}
	\phi^\text{I}_\omega(V)=\int_0^\infty\frac{d\sigma}{{\sqrt{2\pi}}} \left\{\tilde{\phi}^\text{I}_\omega(\sigma) e^{-i\sigma V}-e^{-\pi\omega/a}\tilde{\phi}^\text{I}_\omega(\sigma) e^{i\sigma V}\right\}.
    \label{eq:introunruh1}
\end{align}
Since $e^{-\pi\omega/a}<1$ holds for $\omega>0$, $\phi^\text{I}_\omega(V)$ is a positive norm mode $(\phi_\omega^{\text{I}},\phi_\omega^{\text{I}})>0$ with respect to the Klein-Gordon inner product.\footnote{
In this article, Klein-Gordon inner product of two functions $f,g$ on a constant $U$ surface is defined by
\begin{align}
	(f,g):=-i\int_{-\infty}^{+\infty} dV\{f\partial_V g^*-g^*\partial_V f\}.
\end{align}}
Let us consider the analytic continuation of $\phi^\text{I}_\omega(V)$ for $V>0$ to the negative real line $V<0$ while avoiding the singularity at $V=0$ by a small semi-circle in the lower half plane: 
%%%
\begin{align}
	\phi^{\text{I}\to \text{II}}_\omega(V):&=e^{-\pi\omega/a}\exp\left(-i\frac{\omega}{a}\log(-aV)\right)\theta(-V) \notag \\
	&=e^{-\pi\omega/a}\phi^\text{I}_\omega(-V).
  %  &\equiv e^{-\pi\omega/a}\phi^\text{II}_\omega(V).
\end{align}
%%%
This function has support in the region $V<0$. 
The Fourier decomposition of $\phi^{\text{I}\to \text{II}}_\omega(V)$ is
%%%
\begin{align}
	\phi^{\text{I}\to \text{II}}_\omega(V)=-e^{-\pi\omega/a}\int_0^\infty \frac{d\sigma}{{\sqrt{2\pi}}}\left\{e^{-\pi\omega/a}\tilde{\phi}^\text{I}_\omega(\sigma)e^{-i\sigma V}-\tilde{\phi}^\text{I}_\omega(\sigma)e^{i\sigma V}\right\}.
    \label{eq:introunruh2}
\end{align}
%%%
Since $e^{-\pi\omega/a}<1$ holds for $\omega>0$, $\phi^\text{II}_\omega(V)$ is a negative norm mode, i.e., $(\phi_\omega^{\text{II}},\phi_\omega^{\text{II}})<0$ with respect to the Klein-Gordon inner product.
From Eqs.~\eqref{eq:introunruh1} and \eqref{eq:introunruh2}, we can see that the following combination of the functions contains purely positive frequency contribution with respect to the coordinate $V$:
%%%
\begin{align}
\Phi_\omega(V):&=\phi^\text{I}_\omega(V)+\phi^{\text{I}\to \text{II}}_\omega(V)\\
	&=\phi^\text{I}_\omega(V)+e^{-\pi\omega/a}\phi^{\text{II}}_\omega(V)\\
	&=\phi^\text{I}_\omega(V)+e^{-\pi\omega/a}\left(\phi^{\text{II}}_{-\omega}(V)\right)^*,
\end{align}
%%%
where $\phi^{\text{II}}_\omega(V):=e^{\pi\omega/a}\phi^{\text{I}\to \text{II}}_\omega(V)$. 
$\Phi_\omega(V)$ is called the Unruh mode function defined throughout the whole Minkowski spacetime.
 We use $\phi^{\text{II}}_{-\omega}(V)$ in the last line, since they are positive norm modes. 
 The mode $\phi^\text{I}_\omega$ is  called Rindler mode and the  modee $\phi^\text{II}_{-\omega}$ is called Milne mode.
The mode defined by 
%%%
\begin{align}
	\Psi_{-\omega}(V):=\left(\phi^\text{I}_\omega(V)\right)^*+e^{\pi\omega/a}\phi^{\text{II}}_{-\omega}(V)
\end{align}
%%%
is orthogonal to the mode $\Phi_\omega(V)$ and has a positive norm with respect to the Klein-Gordon inner product.
By normalizing these mode functions, the Bogoliubov transformation between the Rindler/Milne mode functions $(\phi_\omega^\text{I}, \phi_{-\omega}^\text{II})$ and the Unruh mode functions $(\Phi_\omega, \Psi_{-\omega})$ is given by
%%%
\begin{align}
	\Phi_\omega(V)&=\frac{e^{\pi\omega/2a}}{\sqrt{2\sinh\left(\pi\omega/a\right)}}\phi^\text{I}_\omega(V)+\frac{e^{-\pi\omega/2a}}{\sqrt{2\sinh\left(\pi\omega/a\right)}}\left(\phi^{\text{II}}_{-\omega}(V)\right)^*,\\
	\Psi_{-\omega}(V)&=\frac{e^{-\pi\omega/2a}}{\sqrt{2\sinh\left(\pi\omega/a\right)}}\left(\phi^\text{I}_\omega(V)\right)^*+\frac{e^{\pi\omega/2a}}{\sqrt{2\sinh\left(\pi\omega/a\right)}}\phi^{\text{II}}_{-\omega}(V),
\end{align}
where we redefined $\phi^{\text{I},\text{II}}_\omega$ so that they are normalized as
\begin{align}
	\phi^{\text{I},\text{II}}_\omega\to\frac{1}{\sqrt{4\pi|\omega|}}\phi^{\text{I},\text{II}}_\omega.
\end{align}
%%%
By inverting this relation, we obtain
%%%
\begin{align}
	\label{eq:bogrind}
	\phi^\text{I}_\omega(V)&=\frac{e^{\pi\omega/2a}}{\sqrt{2\sinh\left(\pi\omega/a\right)}}\Phi_\omega(V)-\frac{e^{-\pi\omega/2a}}{\sqrt{2\sinh\left(\pi\omega/a\right)}}\left(\Psi_{-\omega}(V)\right)^*,\\
	\label{eq:bogrindpart}
	\phi^{\text{II}}_{-\omega}(V)&=-\frac{e^{-\pi\omega/2a}}{\sqrt{2\sinh\left(\pi\omega/a\right)}}\left(\Phi_\omega(V)\right)^*+\frac{e^{\pi\omega/2a}}{\sqrt{2\sinh\left(\pi\omega/a\right)}}\Psi_{-\omega}(V).
\end{align}
%%%
Now, we introduce annihilation operators by using the Klein-Gordon inner products between field operator $\hat{\phi}$ and mode functions:
\begin{align}
	\hat{a}^\text{I}_\omega&=(\hat{\phi},\phi^\text{I}_\omega),
	\hspace{0.5cm}\hat{a}^\text{II}_\omega=(\hat{\phi},\phi^\text{II}_{-\omega}),\\
	\hat{a}^\Phi_\omega&=(\hat{\phi},\Phi_\omega),
	\hspace{0.5cm}\hat{a}^\Psi_\omega=(\hat{\phi},\Psi_{-\omega}).
\end{align}
%%%
Using \eqref{eq:bogrind} and \eqref{eq:bogrindpart}, these annihilation operators can be related as
\begin{align}
	\label{eq:bogop}
	\hat{a}^\text{I}_\omega&=\frac{e^{\pi\omega/2a}}{\sqrt{2\sinh\left(\pi\omega/a\right)}}\hat{a}^\Phi_\omega+\frac{e^{\pi\omega/2a}}{\sqrt{2\sinh\left(\pi\omega/a\right)}}(\hat{a}^\Psi_\omega)^\dag,\\
	\label{eq:bogpaop}
	\hat{a}^\text{II}_\omega&=\frac{e^{\pi\omega/2a}}{\sqrt{2\sinh\left(\pi\omega/a\right)}}\hat{a}^\Psi_\omega+\frac{e^{\pi\omega/2a}}{\sqrt{2\sinh\left(\pi\omega/a\right)}}(\hat{a}^\Phi_\omega)^\dag.
\end{align}
%%%
From the orthogonality of the mode functions
%%%
\begin{align}
	(\phi^\text{I}_\omega,\phi^\text{II}_{-\omega})&=\left(\phi^\text{I}_\omega,\left(\phi^\text{II}_{-\omega}\right)^*\right)=0,\\
	(\Phi_\omega,\Psi_{-\omega})&=\left(\Phi_\omega,\left(\Psi_{-\omega}\right)^*\right)=0,
\end{align}
 we can show the independency of the annihilation operators:
\begin{align}
	[\hat{a}^{\text{I}}_\omega,\hat{a}^{\text{II}}_\omega]=[\hat{a}^{\text{I}}_\omega,\left(\hat{a}^{\text{II}}_\omega\right)^\dag]=0,\\
	[\hat{a}^\Phi_\omega,\hat{a}^{\Psi}_\omega]=[\hat{a}^{\Phi}_\omega,\left(\hat{a}^{\Psi}_\omega\right)^\dag]=0.
\end{align}
%%%
The independence of annihilation operators implies the existence of two different particle modes. 
Since there are two pairs $\{\hat{a}^\text{I}_\omega,\hat{a}^\text{II}_\omega\}$ and $\{\hat{a}^\Phi_\omega,\hat{a}^\Psi_\omega\}$ of independent annihilation operators, we have two different vacuum states $|00\rangle_{\text{I II}}$  and $|00\rangle_{\Phi\Psi}$ satisfying
\begin{align}
	0&=\hat{a}_\omega^\text{I}|00\rangle_{\text{I II}}=\hat{a}_\omega^\text{II}|00\rangle_{\text{I II}},\\
	0&=\hat{a}_\omega^\Phi|00\rangle_{\Phi\Psi}=\hat{a}_\omega^\Psi|00\rangle_{\Phi\Psi}.
\end{align}
These vacuum states are related as
\begin{align}
	|00\rangle_{\Phi\Psi}=\sum_{n=0}^\infty\frac{e^{-n\pi\omega/a} }{\sqrt{1-e^{-2\pi\omega/a}}}|nn\rangle_{\text{I II}}.
\end{align}
%%%
Tracing out $\text{II}$ degrees of freedom from the vacuum state $|00\rangle_{\Phi\Psi}$, we have the mixed density matrix
%%%
\begin{align}
	\hat{\rho}_\text{I}=\sum_{n=0}^\infty\frac{e^{-2n\pi\omega/a} }{1-e^{-2\pi\omega/a}}|n\rangle\langle n|_\text{I}.
\end{align}
%%%
In contrast, this mixed state for $\text{I}$ can be purified by $\text{II}$ degrees of freedom to produce the pure vacuum state $|00\rangle_{\Phi\Psi}$. In this sense, the mode $\hat{a}_\text{II}$ that appears as the counterpart of the Bogoliubov transformation is called the partner mode of the mode $\hat{a}_\text{I}$.
Indeed, Eq.~(\ref{eq:bogpaop}) is a one example of the partner formula given by Hotta-Sch\"utzhold-Unruh \cite{hotta2015partner}. See Appendix~\ref{sec:revpartner} for a general discussion of the partner formula.

%%%%%%%%%%%%%%%%%%%%%%%%%%%%%%%%%%%%%%%%%%%%%%%%%
%%%%%%%%%%%%%%%%%%%%%%%%%%%%%%%%%%%%%%%%%%%%%%%%%%
\subsection{Profile of the Rindler Mode and Its Partner}\label{sec:visurindler}
\begin{figure}[H]
	\centering
        \includegraphics[width=\linewidth]{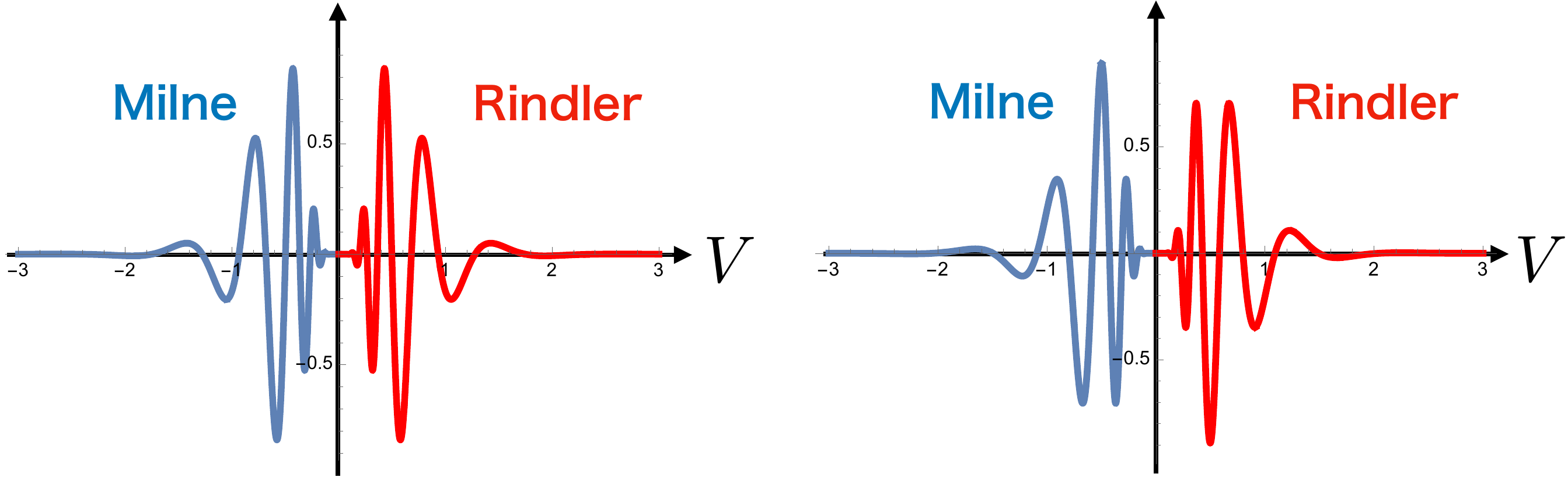}
        \caption{Left panel:  Profiles of the Rindler mode $q_D(V)$  and  its partner $q_P(V)$ (we call it the Milne mode). They are symmetric about the vertical axis. Right panel: Profiles of Rindler mode $p_D(V)$ and its partner  $p_P(V)$. They are symmetric about the origin. We have chosen the Gaussian weighting function $F(\omega)\propto e^{-10(\omega-1)^2}$ for this plot.} 
         \label{fig:partner}
\end{figure}

By superposing the Rindler mode function, we can construct the detector mode of the Rindler observer:
\begin{align}
	\varphi_D(V):=\int_0^\infty\dd\omega F(\omega)\phi^\text{I}_\omega(V),
\end{align}
where $F(\omega)$ is a weighting function. Using the representation of the field operator with the left-moving mode
%%%
\begin{equation}
\hat\phi(V)=\int_0^\infty d\omega(\hat a_\omega^\text{I}\,\phi_\omega^\text{I}(V)+\text{h.c.}),
\end{equation}
%%%
and its canonical conjugate operator $\hat\Pi(V):=\hat\phi'(V)$, the annihilation operator associated with the detector mode is defined by the Klein-Gordon inner product with the field operator 
\begin{align}
	\hat{a}_D&:=(\hat{\phi}(V),\varphi_D(V))\notag\\
	&=\int_0^\infty\dd\omega F(\omega)(\hat{\phi}(V),\phi^\text{I}_\omega(V))\notag\\
	&=\int_0^\infty\dd\omega F(\omega)\hat{a}^\text{I}_\omega,
	\label{eq:annidet}
\end{align}
where $F(\omega)$ should satisfy the  normalization condition
\begin{align}
	1=[\hat{a}_D,\hat{a}_D^\dag]=\int_0^\infty\dd\omega |F(\omega)|^2.
\end{align}
We can relate the superposition of the mode functions to the spatial profile of  the detector mode. 
We define the canonical pair for the detector mode by
%%%
\begin{align}
	\hat{Q}_D=\frac{\hat{a}_D+\hat{a}_D^\dag}{\sqrt{2}},\quad \hat{P}_D=\frac{\hat{a}_D-\hat{a}_D^\dag}{\sqrt{2}i},
 \quad [\hat Q_D,\hat P_D]=i.
	\label{eq:qpdet}
\end{align}
%%%
These local operators defined from the quantum field $\hat\phi(V)$ can be expressed using spatial profile functions $q_D(y)$ and $p_D(y)$ as follows:\footnote{For a chiral scalar field, we use a gauge invariant field operator $\hat\Pi(V):=\pa_V\hat\phi(V)$ to define a local operator $\hat Q_P(V)=\int dV q_P(V)\hat\Pi(V)$ where $q_P(V)$ is a profile function for the local operator.}
\begin{align}
	\hat{Q}_D&=\int_{-\infty}^\infty\dd V q_D(V)\int_0^\infty\dd\omega \left\{\pa_t\phi^\text{I}_\omega(V)\hat{a}^\text{I}_\omega+\text{h.c.}\right\},\\
	\hat{P}_D&=\int_{-\infty}^\infty \dd V p_D(V)\int_0^\infty\dd\omega\left\{\pa_t\phi^\text{I}_\omega(V)\hat{a}^\text{I}_\omega+\text{h.c.}\right\}.
\end{align} 
%%%
By using Eq.~(\ref{eq:annidet}) and Eq.~(\ref{eq:qpdet}), profile functions $q_D(y), p_D(y)$ and the weighting function  $F(\omega)$ are related as
\begin{align}
	\int\dd V q_D(V)\pa_t\phi^\text{I}_\omega(V)&=\frac{F(\omega)}{\sqrt{2}},\quad
	\int\dd V p_D(V)\pa_t\phi_\omega^\text{I}(V)=\frac{F(\omega)}{\sqrt{2}i}.
\end{align}
%%%
The canonical commutation relation for the left-moving mode is
%%%
\begin{align}
	\left[\hat{\phi}(V),\hat{\Pi}(V')\right] 
	&=\int_0^\infty\dd\omega \left\{\phi^\text{I}_\omega(V)\left(\pa_t\phi^\text{I}_\omega(V')\right)^*-\pa_t\phi^\text{I}_\omega(V)\left(\phi^\text{I}_\omega(V')\right)^*\right\}\nn
	&\hspace{1cm}+\int_0^\infty\dd\omega \left\{\phi^{\text{II}}_{-\omega}(V)\left(\pa_t\phi^{\text{II}}_{-\omega}(V')\right)^*-\pa_t\phi^{\text{II}}_{-\omega}(V)\left(\phi^{\text{II}}_{-\omega}(V')\right)^*\right\}\\
 &\equiv\frac{i}{2}\delta\left(V-V'\right). \notag
\end{align}
%%%
Since the mode functions $\phi^\text{I}_\omega$ and $\phi_{-\omega}^{\text{II}}$ are restricted to  $V>0$ and $V<0$, respectively, the following normalization conditions hold:
\begin{align}
	\frac{i}{2}\delta(V-V')&=\int_0^\infty\dd\omega \left\{\phi^\text{I}_\omega(V)\left(\pa_t\phi^\text{I}_\omega(V')\right)^*-\pa_t\phi^\text{I}_\omega(V)\left(\phi^\text{I}_\omega(V')\right)^*\right\}\quad (V,V'>0),\\
	\frac{i}{2}\delta(V-V')&=\int_0^\infty\dd\omega \left\{\phi^{\text{II}}_{-\omega}(V)\left(\pa_t\phi^{\text{II}}_{-\omega}(V')\right)^*-\pa_t\phi^{\text{II}}_{-\omega}(V)\left(\phi^{\text{II}}_{-\omega}(V')\right)^*\right\}\quad (V,V'<0).
\end{align}
%%%
By using these identities we can express the spatial profile of the detector mode in terms of the weighting function of the detector mode:
%%%
\begin{align}
	q_D(V)&=-\sqrt{2}i\int_0^\infty\dd\omega\left\{F(\omega)\left(\phi^\text{I}_\omega(V)\right)^*-F^*(\omega)\phi^\text{I}_\omega(V)\right\}\notag\\
	&=2\sqrt{2} \Im\left[\int_0^\infty\dd\omega F(\omega)\left(\phi^\text{I}_\omega(V)\right)^*\right],\\
	p_D(V)&=-\sqrt{2}\int_0^\infty\dd\omega\left\{F(\omega)\left(\phi^\text{I}_\omega(V)\right)^*+F^*(\omega)\phi^\text{I}_\omega(V)\right\}\notag\\
	&=-2\sqrt{2} \Re\left[\int_0^\infty\dd\omega F(\omega)\left(\phi^\text{I}_\omega(V)\right)^*\right].
\end{align}
%%%
The profile for the partner mode  $\hat{a}_P$ is slightly complicated in general. However, if the weighting  function $F(\omega)$ of the detector mode has a sharp peak at some $\omega$, we can approximate the partner mode as 
\begin{align}
	\hat{a}_P\approx\int_0^\infty\dd\omega F(\omega)\hat{a}^{\text{II}}_\omega,
	\label{eq:singlemodeapprox}
\end{align}
%%%%
where $\hat a_\omega^\text{II}$ is an annihilation operator associated with the Rindler mode $\phi_{-\omega}^\text{II}$. See Appendix~\ref{sec:ap-B} for the derivation 
%\anno{[I am not sure Appendix B explains the derivation of this equation (YN)]}
of this approximation.
Then the partner mode is expressed using its profile function $q_P(y)$ and $p_P(y)$ as
\begin{align}
	\hat{Q}_P&=\int \dd V q_P(V)\int_0^\infty\dd\omega\left\{\pa_t\phi^{\text{II}}_{-\omega}(V)\hat{a}^{\text{II}}_{\omega}+\text{h.c.}\right\},\\
	\hat{P}_P&=\int \dd V p_P(V)\int_0^\infty\dd\omega\left\{\pa_t\phi^{\text{II}}_{-\omega}(V)\hat{a}^{\text{II}}_{\omega}+\text{h.c.}\right\},
\end{align}
%%%
with 
%%%
\begin{align}
	q_P(V)&=2\sqrt{2} \Im\left[\int_0^\infty\dd\omega F(\omega)\left(\phi^{\text{II}}_{-\omega}(V)\right)^*\right],\\
	p_P(V)&=-2\sqrt{2} \Re\left[\int_0^\infty\dd\omega F(\omega)\left(\phi^{\text{II}}_{-\omega}(V)\right)^*\right].
\end{align}
We show the profiles of the Rindler mode and its partner mode  in Fig.~\ref{fig:partner}. These profiles correspond to the in-vacuum state at the past null infinity of the moving mirror model presented in the next section.  They are mirror-reversed images about $V=0$ and they do not have any overlap as expected.
%%%%%%%%%%%%%%%%%%%%%%%%%%%%%%%%%%%%%%%
%%%%%%%%%%%%%%%%%%%%%%%%%%%%%%%%%%%%%%%%%%%%%%%%%%%
%%%%%%%%%%%%%%%%%%%%%%%%%%%%%%%%%%%%%%%%%%%%%%%%%%%

\section {Partner in Moving Mirror Model and vacuum fluctuation scenario}\label{sec:partner moving mirror}
The moving mirror model involves a massless scalar field in the 1+1-dimensional Minkowski spacetime. The scalar field is subject to a Dirichlet boundary at the  perfectly reflecting mirror. We define the worldline of the mirror as $x=z(t)$.
 Therefore, we consider the scalar field
with the boundary condition ${\phi}(t,z(t))=0$. 
Using the null coordinates $(U,V)$,
 the general solution of the scalar field is
 %%%
\begin{align}
	\phi(U,V)=f(U)+g(V),
	\label{eq:solmov}
\end{align}
%%%
where $f,g$ are arbitrary functions. Note that the right-moving solution and the left-moving solution decouple.
 To ensure that our solution satisfies the boundary condition, we express the boundary in the null coordinates. A constant $U$ line intersects with the worldline of the mirror at a single point, which can be represented as $(\tau_U,z(\tau_U))$ satisfying
$\tau_U-z(\tau_U)=U$.
Thus, the $V$ coordinate of the intersection point for a given $U$ is given by 
%%%
\begin{align}
	V_U=\tau_U+z(\tau_U)\equiv p(U),
\end{align}
%%%
and we refer to $p(U)$ as the ray tracing function.
The boundary condition can then be rewritten as $\phi(U,p(U))=0$. 
Consequently, arbitrary functions in Eq.~(\ref{eq:solmov}) must be related by means of equation $f(U)=-g\left( p(U)\right)$.
Thus, the solution of the wave equation with the required boundary condition is
%%%
\begin{align}
	\phi(U,V)=-g(p(U))+g(V).
\end{align}
%%%
The left-moving wave  $g(V)$ is reflected in the right-moving wave $g(p(U))$ at the moving
boundary $V=p(U)$.

\begin{figure}[H]
	\begin{center}
        \includegraphics[width=9cm]{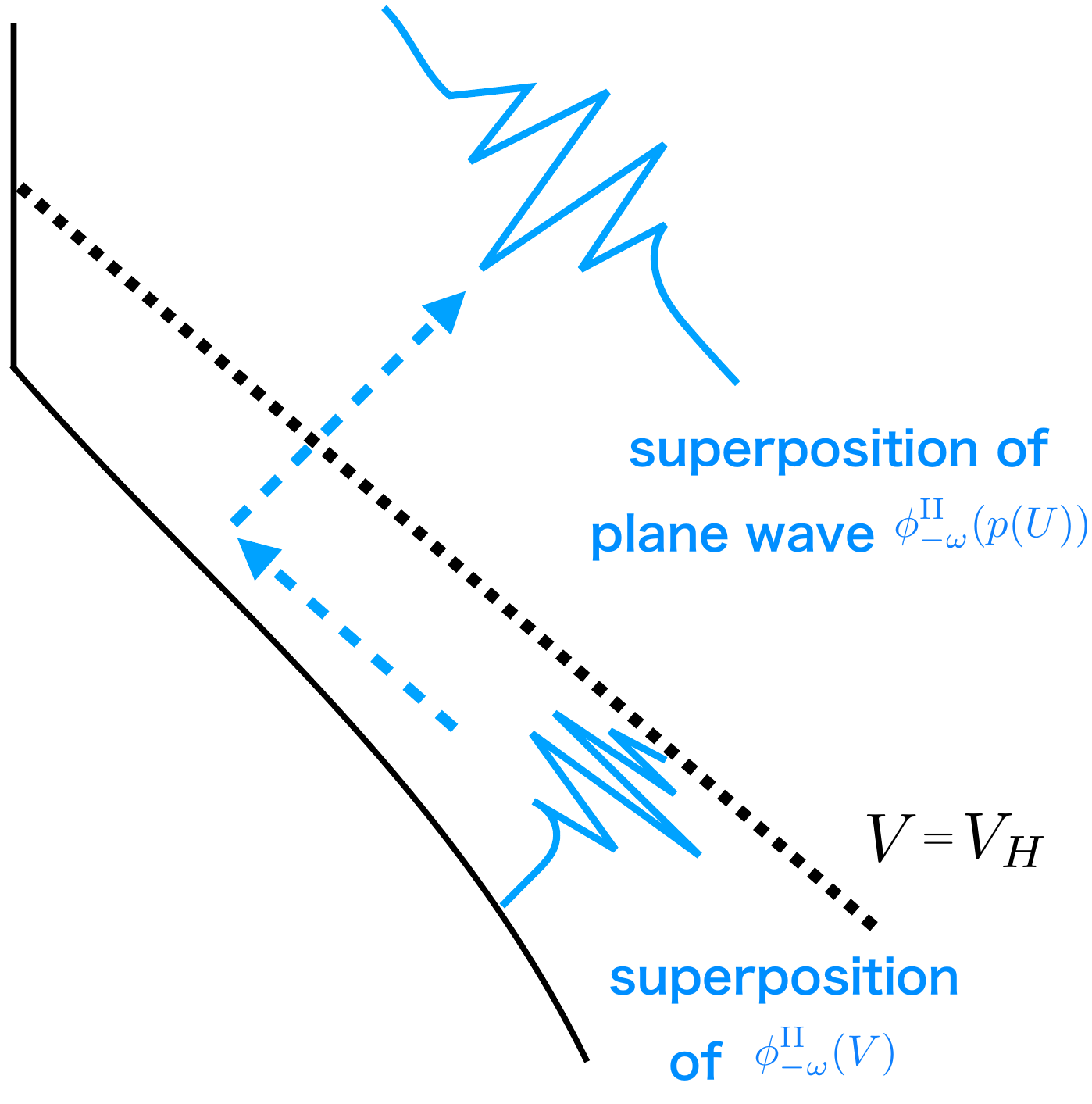}
        \caption{Relation between modes at the future null infinity and at the past null infinity.}
        \label{fig:modestr}
        \end{center}
\end{figure}

For a mirror trajectory which asymptotically approaches the null line $V=0$, the ray-tracing function is given by 
%%%
\begin{align}
	p(U)=-\frac{1}{a}e^{-a U},
	\label{eq:rtnull}
\end{align}
%%%
where $a$ is the proper acceleration of the mirror.
By reflection at the moving mirror, the left-moving Milne mode $\phi^{\text{II}}_{-\omega}(V)$ becomes the following right-moving wave:
%%%
\begin{align}
	\frac{1}{\sqrt{4\pi\omega}}e^{-i\omega U}=\phi^{\text{II}}_{-\omega}(p(U)).
\end{align}
%%%
Therefore, Milne particles  appear if we measure the plane wave-type Minkowski mode at the future null infinity $V=\infty$, assuming that the mirror trajectory has the portion that approaches a null line asymptotically (Fig.~\ref{fig:modestr}).
For a suddenly stopping moving mirror trajectory, considered by R.Wald \cite{wald2019particle}, the trajectory is given by 
\begin{align}
	z(t)=\begin{cases}
	0 &\quad (t<0)  \\
	-t-e^{-2\kappa t}/{2\kappa}+{1}/{2\kappa}&\quad (0\le t\le t_*-\varepsilon)\\
	z(t_*)&\quad(t\ge t_*+\varepsilon).
	\end{cases}
	\label{eq:mirtrajwald}
\end{align}
Here, we determine the values of $z(t_*)$ and $z'(t_*)$ based on the behavior of $z$ when $z\le z_*$.
$\varepsilon$ is a small parameter, and when we perform numerical simulations, $z(t)$ in $t<t_*-\varepsilon$ and $z(t)$ in $t>t_*+\varepsilon$ are smoothly connected by interpolation.
The ray tracing function $p(U)$ for this trajectory is given by
\begin{align}
	p(U)=\left\{\begin{array}{l}
	U\hspace{0.5cm}(U\le0)\\
	W(-e^{-\kappa U-1/2}/2)/\kappa+1/2\kappa\hspace{0.5cm}(0\le U\le t_*-z(t_*)-\varepsilon)\\
	U+2z(t_*)\hspace{0.5cm}(U\ge t_*-z(t_*)+\varepsilon).
	\end{array}\right.,
\end{align}
where $W(x)$ is the Lambert W function defined as the solution $y$ to the equation $ye^y=x$.
This trajectory has an asymptotic null line $V=1/2\kappa$ for $U\gg1$ and the ray tracing function is approximated by $p(U)\sim e^{-\kappa U}/2\kappa+1/2\kappa$ in this region.
Since this type of mirror trajectory experiences a large deceleration for $t_*-z(t_*)-\varepsilon\le u\le t_*-z(t_*)+\varepsilon$, a huge amount of energy flux (burst) is emitted in this region.
Following the discussion in the previous sections (Eq.~(\ref{eq:singlemodeapprox}) of Sec.~\ref{sec:visurindler}), we know that the partner mode for the superposition of $\phi^{\text{II}}_{-\omega}$ is the superposition of $\phi^{\text{I}}_{\omega}$ with the same superposition coefficients. To obtain the partner mode at the future null infinity, all we have to consider is the image of modes reflected by the moving mirror.
%%%%%%%%%%%%%%%%%%%%%%%%%%%%%%%
\begin{figure}[H]
	\begin{center}
        \includegraphics[width=\linewidth]{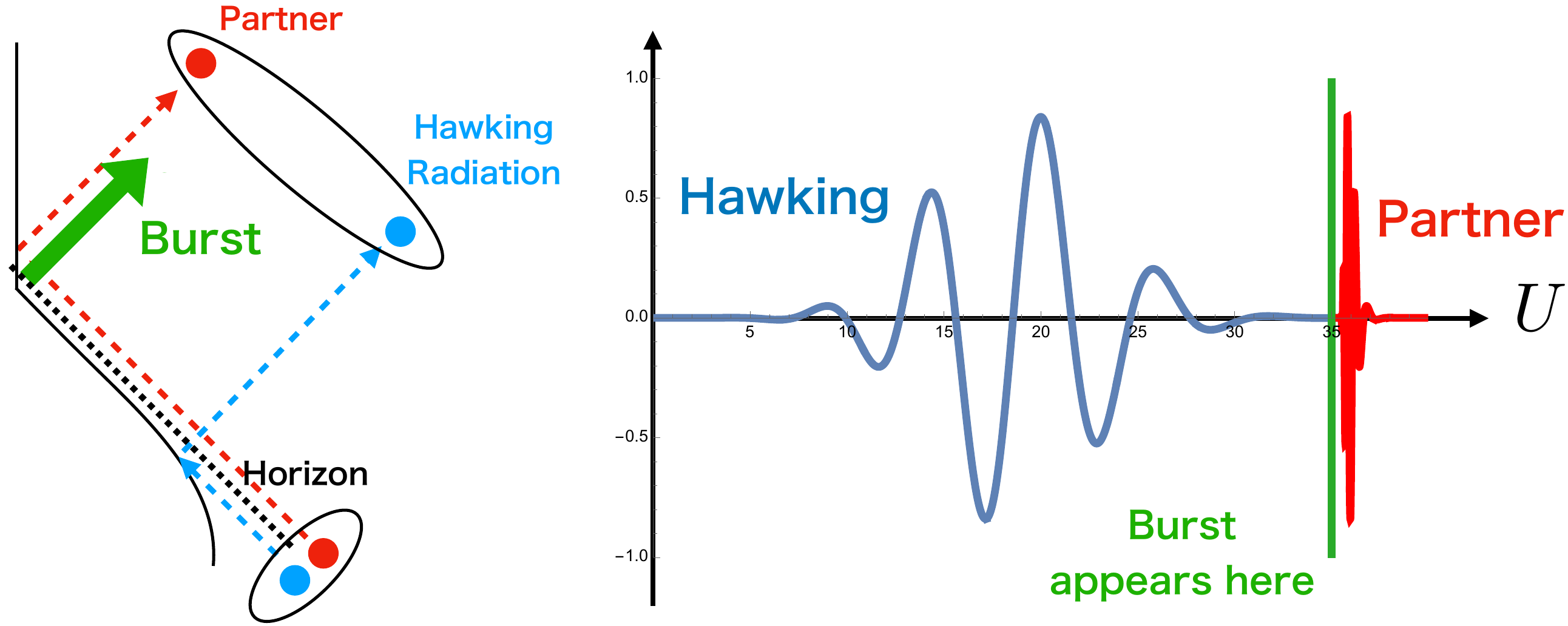}
        \caption{Mode structures for the suddenly stopping mirror case. Schematic picture of Hawking particles and partner particles (left panel). Shape of the partner mode $q_P(U)$(red) for a given Hawking mode $q_D(U)$(blue)(right panel). The green line denotes the location where the burst appears. We have chosen the weighting function $F(\omega)\propto e^{-10(\omega-1)^2}$.}
        \label{fig:prtnrwald}
        \end{center}
\end{figure}
%%%%%%%%%%%%%%%%%%%%%%%%%%%%%%%%
A schematic diagram of Hawking mode, its partner and the burst is depicted in the left panel of Fig.~\ref{fig:prtnrwald}. 
The profiles of the Hawking mode and partner mode (smearing functions $q_D(U)$ and $q_P(U)$) at the future null infinity are also plotted in the right panel of Fig.~\ref{fig:prtnrwald}.
At the past null infinity, the partner mode and the Hawking mode appear on opposite sides of the past Rindler horizon. At the future null infinity, they appear on opposite sides of the null line $U=1/2\kappa-2z(t_*)$ that intersects at the point where the horizon and the mirror trajectory cross.
 It is worth noting that the burst appears on the Hawking mode sides (i.e. $U\le1/2\kappa-2z(t_*)$) at the future null infinity and it can be regarded as late-time Hawking radiation. This is because the horizon and the worldline of the mirror intersect after the mirror stops.
 We can estimate the energy of the emitted partner mode by the expectation value $E_P:=\langle\hat{a}_P{}^{\!\dag}\hat{a}_P+1/2\rangle_{in}$, and this can be evaluated as (see Appendix~\ref{ap-a}) for a detailed calculation)
 %%%
 \begin{align}
 	E_P=\frac{1}{2}\int dU \dd {U'} \left\{\langle\hat{T}_{UU}\rangle_{ren}+\frac{1}{(U-U'-i\varepsilon)^2}\right\}(q_P(U)q_P(U')+p_P(U)p_P(U'))+\frac{1}{2}.
 \end{align} 
 %%%
 The first term in the curly bracket is the expectation value of the renormalized energy-momentum tensor, which corresponds to the energy flux radiated from the mirror, and the other terms correspond to the contributions of vacuum fluctuations. As we have obsereved in the left panel in Fig. \ref{fig:prtnrwald}, %the {\color{blue}support?} of $\langle\hat{T}_{UU}\rangle_{ren}$ and the profile of the partner mode do not overlap. 
 the energy momentum tensor $\langle\hat{T}_{UU}\rangle_{ren}$ takes non-zero value only on the Hawking mode side ($U\le1/2\kappa-2 z(t_*)$).
 Thus the renormalized energy momentum tensor does not contribute to $E_P$.
 Therefore, the cost of energy to emit the partner can only be contribution from the vacuum fluctuations.
% Fig.~\ref{fig:prtnrwald}, we plotted the shape of the smearing functions $q_D(u)$ and $q_P(u)$ which corresponds to the shape of the Hawking mode and partner mode in the future null infinity.
%We can see that the partner mode appears after the shock wave appeared as we have expected. Nevertheless we can check that the partner mode and the shock wave are the different one, still, there are possibility that the appearance of the shock wave is necessary for returning partner of the early time Hawking radiation into the spacetime. This possibility corresponds to the burst scenario.

%%%%%%%%%%%%%%%%%%%%%%%%%%%%%%%%%%%%%%%%%%%%%%%%%%%%%%%%%%%%%%%%%%%%%%%%%%%%%%%%%%%%%%%%%%%%%%%%%%%%%%%%%%%%%%
\section {Wald's argument and its deficiency}\label{sec:wald}
\subsection{Consideration by Wald}
In this section, we briefly review the argument of R. Wald on Hawking particles and their partners \cite{wald2019particle}. He considered that the emission of the burst can be explained by entanglement of Hawking radiation and vacuum fluctuations. 
If we accept his statement, to recover information, the emission of the burst is necessary to return the partner of Hawking radiation, and the energy of the burst should be regarded as an indirect energy cost of purification.
%%%%%%%%%%%%%%%%%%%%%%%%%%%%%%%
\begin{figure}[H]
	\begin{center}
        \includegraphics[width=7cm]{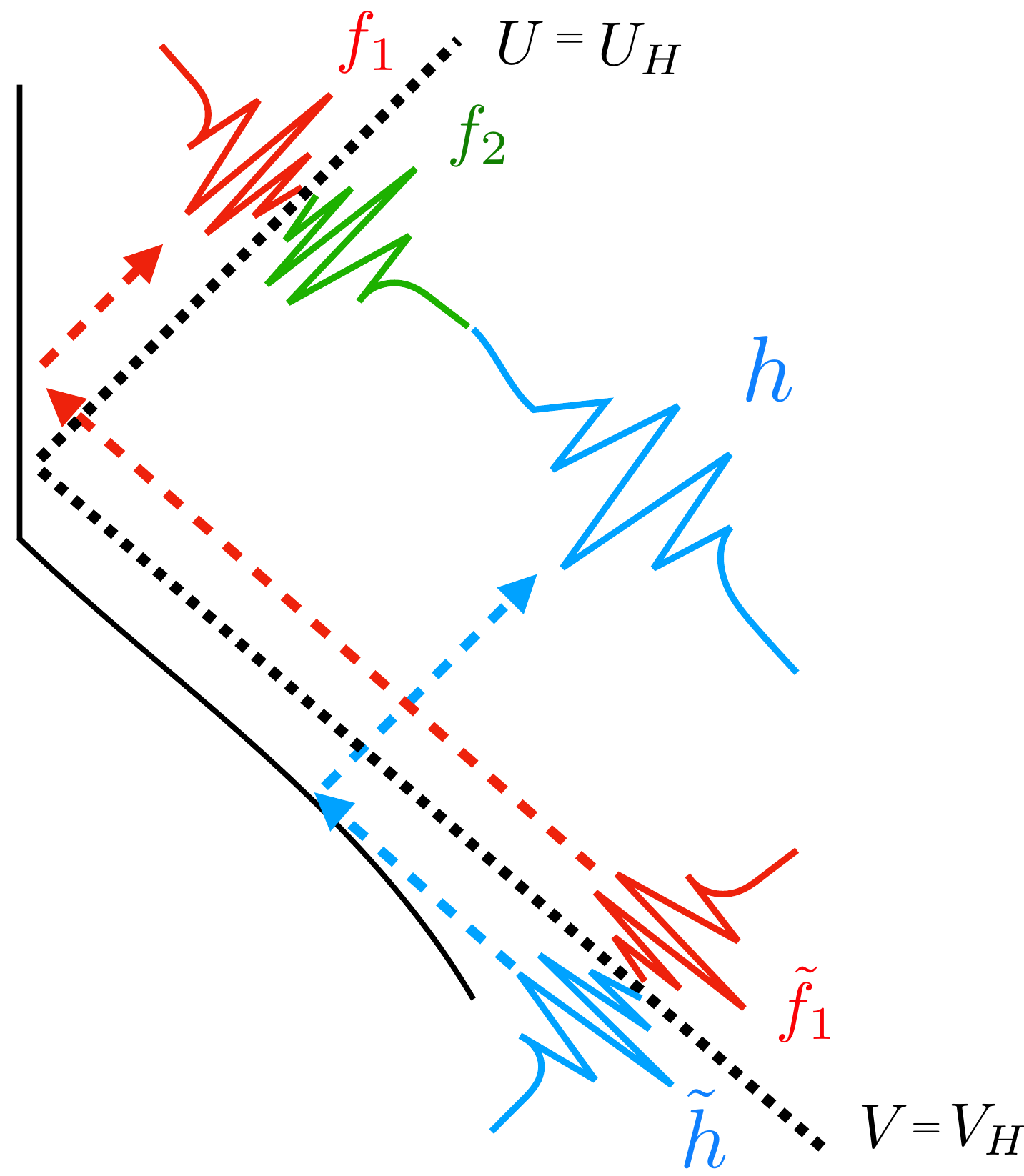}
        \caption{Schematic picture of modes of the suddenly stopping mirror}
        \label{fig:waldimage}
        \end{center}
\end{figure}
%%%%%%%%%%%%%%%%%%%%%%%%%%%%%%%%
Let us focus on the mirror trajectory Eq.~(\ref{eq:mirtrajwald}) described in the previous section for a moment. We prepare a wave packet $h$, which consists of plane waves at the future null infinity that correspond to Hawking radiation.
By propagating the wave packet backward in time, it is reflected off by the mirror during the mirror's accelerating phase, and then further propagating it backward, we obtain the wave packet $\tilde{h}$, which consists of Milne waves at the past null infinity.
As discussed in Sec.~\ref{sec:rindlerpartner}, the partner of this Milne wave packet $\tilde{h}$ is the Rindler wave packet $\tilde{f}_1$, which is the flipped image of $\tilde{h}$ with respect to the asymptotic line $V=V_H$. Taking into account the evolution of the wave packet $\tilde{f}_1$, we obtain the partner wave packet $f_1$ of Hawking radiation $h$ at the future null infinity.

Up to this point, there is no deviation from the standard vacuum fluctuation scenario discussed in the previous sections. Wald's consideration is as follows:
the partner wave packet $f_1$ consists of Milne waves, since it is obtained by reflection off the mirror at rest. Since the Milne wave packet can be purified by the Rindler wave packet, the Rindler wave packet $f_2$ also purifies the Milne wave packet $f_1$. The Rindler wave packet $f_2$ is obtained by flipping $f_1$ at $U=U_H$, corresponding to the coordinate $U$ of the intersecting point of $V=V_H$ and the mirror's worldline.
However, since the vacuum fluctuation $f_1$ is already entangled with Hawking radiation $h$, $f_2$ cannot be in a state that is also entangled with $f_1$ according to the monogamy property of quantum entanglement.
Therefore, we conclude that $f_2$ should not be in the state of vacuum fluctuation entangled with $f_1$, but rather be real particles.
The energy cost of these real particles can be estimated by the typical Milne frequency of the Milne wave packet $f_1$ or $f_2$. The Milne wave packet $f_1$ or $f_2$ is strongly blueshifted compared to the wave packet $h$, with an extremely high Milne frequency compared to the typical frequency of Hawking radiation and with a point-like  support. Therefore, the  radiation is burst wavelike, with energy that easily exceeds the Planckian mass scale.
 Since the emission of the burst is necessary to return the partner of Hawking radiation, the energy of the burst should be considered as the indirect cost of purification.

The natural extension of Wald's consideration is what happens if we consider a long-propagating mirror, which is analogous to the remnant scenario \cite{good2015signatures,chen2017entropy}. 
By considering a mirror trajectory with a mild deceleration phase, or a mirror trajectory without a deceleration phase, we will obtain a long propagating mirror trajectory.
For this long-propagating mirror trajectory, the partner wave packet $f_1$ is not strongly blue-shifted compared to Hawking radiation $h$. Therefore, the real particle radiation is not burst-like, and its energy does not exceed the Planckian mass energy.   
Wald has already prepared the answer to this question. For a long-propagating mirror trajectory, the width of the partner wave packet is much larger than the Planck mass scale, due to the strong redshift by the mirror reflection. However, if we consider a real black hole in the semiclassical evaporation scenario, the returned partner wave packet must have a width which is smaller than the Planck mass scale because of causality.
Consequently, the long-propagating  mirror trajectory would not adequately describe the real black hole evaporation process. Thus, the mirror trajectory should be of sudden stopping type. However, for this trajectory, we encounter the problem of an indirect energy cost for purification. Consequently, the vacuum fluctuation scenario should be associated with the burst with a huge amount of energy.   

%%%%%%%%%%%%%%%%%%%%%%%%%%%%%%%%%%%%%%%%%%%%%%%%%%%%%%%
%%%%%%%%%%%%%%%%%%%%%%%%%%%%%%%%%%%%%%%%%%%%%%%%%%%%%%%
\subsection{Critiques on Wald's consideration}\label{sec:comwald}
In the previous section, we discussed Wald's examination on the vacuum fluctuation scenario.
The central aspect of Wald's analysis is the connection between the burst and the entanglement among vacuum fluctuations. Consequently, the burst must be viewed as an indirect energy cost of the purification process because of this relation.
By examining the outcome illustrated in Fig.~\ref{fig:prtnrwald} of Sec.~\ref{sec:partner moving mirror}, an overlap is evident between the burst's location and the profile of the wave packet linked to $f_2$ from Sec.~\ref{sec:wald}, thereby supporting Wald's argument.
To further investigate the relationship between the burst and the entanglement among vacuum fluctuations, we examined the following example of a long-propagating  mirror trajectory:
\begin{align}
	z(t)=
    \begin{cases}
	0&\quad(t\le0)\\
	-t-e^{-2\kappa t}/{2\kappa}+{1}/{2\kappa}&\quad(0\le t\le t_*)\\
	z'(t_*)(t-t_*)+z(t_*)&\quad(t_*\le t\le t_{**})\\
	z(t_{**})&\quad(t\ge t_{**}),
	\end{cases}
	\label{eq:mirtra}
\end{align}
For this trajectory, we calculate $z(t_*)$ and $z'(t_*)$ according to the behavior of $z$ when $t\le t_*$, and we ascertain $z(t_{**})$ based on the behavior of $z$ when $t_*\le t\le t_{**}$.
Without the final static phase, it is nothing but a long-propagating mirror that ends with uniform motion. According to causality, we cannot distinguish them from the behavior of Hawking radiation and its partner, as long as the mirror undergoes a sufficiently long period of uniform motion.
The only difference lies in the emission of the burst during the transition from uniform to static motion.
The ray tracing function for this trajectory is as follows
\begin{equation}
	p(U)=\begin{cases}
	U&\quad(U\le0)\\
	W(-e^{-\kappa U-1/2}/2)/\kappa+1/2\kappa&\quad(0\le U\le t_*-z(t_*))\\
	2(U+z(t_*)-z'(t_*)t_*)/(1-z'(t_*))-U&\quad(t_*-z(t_*)\le U\le t_{**}-z(t_{**}))\\
	U+2z(t_{**})&\quad(t_{**}-z(t_{**})\ge t).
	\end{cases}.
\end{equation}
In the numerical simulation, the interpolation functions are incorporated around $t=0$, $t=t_*$, and $t=t_{**}$ to ensure the smoothness of $z(t)$ and $p(U)$.
%%%%%%%%%%%%%%%%%%%%%%%%%%%%%%%
\begin{figure}[H]
	\begin{center}
        \includegraphics[width=\linewidth]{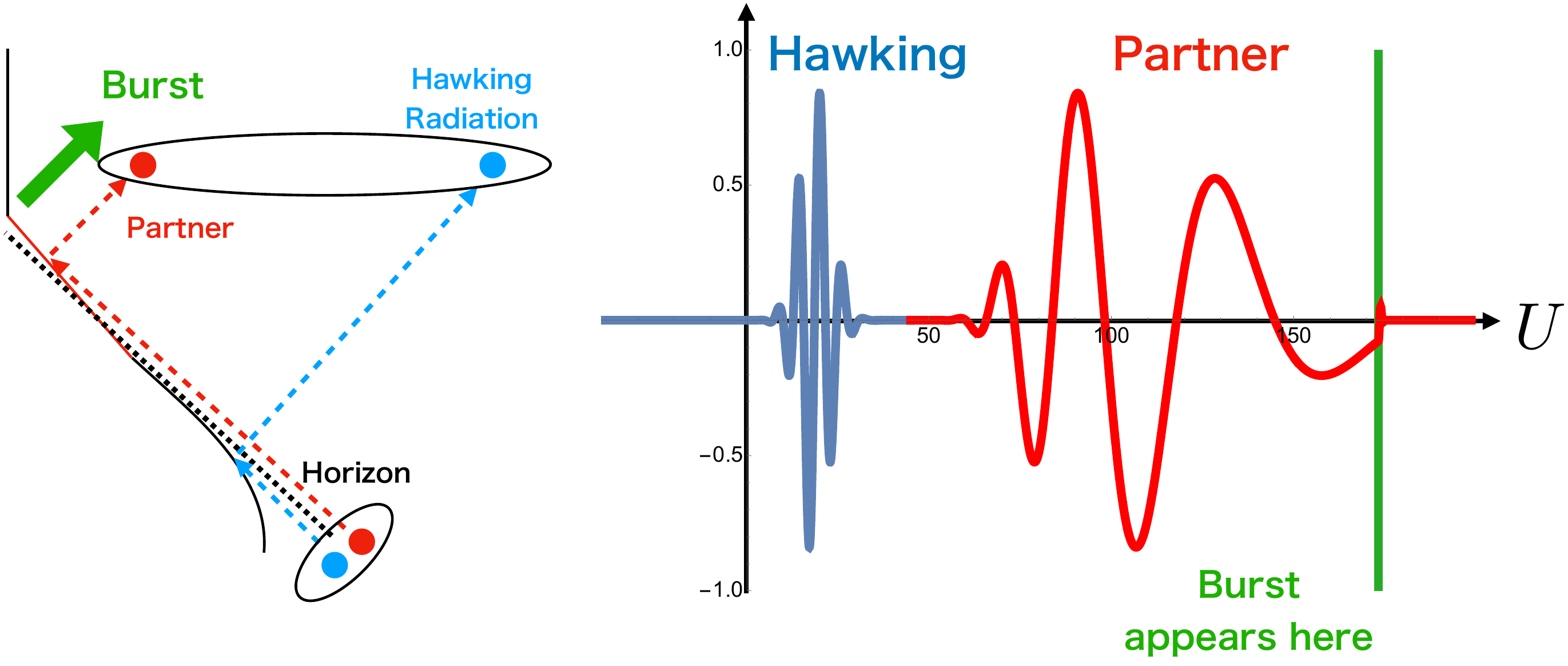}
        \caption{Mode structures for the long-propagating mirror. Schematic picture of the Hawking particle and the partner particle(left panel). Shape of the partner mode $q_P(U)$(red) for given Hawking mode $q_D(U)$(blue)(right panel). The green line corresponds to the location of the burst. We have chosen the weighting function $F(\omega)\propto e^{-10(\omega-1)^2}$.}
        \label{fig:prtnrour}
        \end{center}
\end{figure}
%%%%%%%%%%%%%%%%%%%%%%%%%%%%%%%%
In the left panel of Fig.~\ref{fig:prtnrour}, we present a schematic representation of Hawking radiation, its partner, and the burst. Additionally, the right panel of Fig.~\ref{fig:prtnrour} presents profiles of the Hawking mode and the partner mode (smearing functions $q_D(U)$ and $q_P(U)$) at the future null infinity.
At past null infinity, the partner mode and the Hawking mode lie on opposite sides of the horizon, and at the future null infinity, they also lie on opposite sides of $U=U_H\equiv1/2\kappa-2z(t_*)$ where the horizon and mirror trajectory cross.
 Note that the burst emerges on the side of the partner mode at the future null infinity. This happens because the horizon and worldline of the mirror intersect during the mirror's uniform motion phase. 
 It should be noted that there is no overlap between the profile of the flipped image of the partner of Hawking radiation at $u=u_H$ and the location of burst. Furthermore, assuming that the mirror trajectory undergoes a sufficiently long period of uniform motion, the overlap between the profile of the partner of Hawking radiation and the burst becomes negligible.
If the burst originates from vacuum fluctuation $f_2$, which is entangled with the partner of Hawking radiation as proposed by Wald, there should be an overlap between the location of the burst wave and the profile of vacuum fluctuation $f_2$. However, our example demonstrates that, in general, there is no overlap between the location of the burst and the support of the vacuum fluctuation $f_2$.
Thus, the relationship between the burst and vacuum fluctuations is not as Wald described, and there is no need to consider the energy of the burst as an indirect energy cost of purification.
Before concluding this section, let us briefly explore the entanglement monogamy between Hawking radiation $h$, its partner $f_1$, and the flipped image $f_2$ of $f_1$.
It is not always possible to uniquely determine the purification partner for some mixed quantum states.
%The monogamy issue need not be a concern even if we inadvertently discover two ways to purify a mixed quantum state, as long as the purified total systems do not coincide.
In the case of the moving mirror, purifying $f_1$ with $h$ results in the in-vacuum state, whereas purifying $f_1$ with $f_2$ results in the out-vacuum state.
If the in-vacuum state and the out-vacuum state coincide, then $h$ and $f_2$ must be identical, or one of them must abandon quantum entanglement with $f_1$ due to the monogamy relation.
However, this is not the case for the moving mirror with an accelerating phase, and $f_2$ is still in the state of the vacuum fluctuation.

\bigskip

\section{Conclusion}\label{sec:conclusion}
We have reconsidered Wald's critique on the vacuum fluctuation scenario for black hole evaporation, particularly focused on his argument relating to the burst scenario.
 Wald argues that the emission of the burst at the final stage of black hole evaporation can be explained from the viewpoint of monogamy, suggesting that the burst emission is inevitable for the purification of Hawking radiation with vacuum fluctuations.
 We have shown that there is no overlap between the location of the flipped profile of the partner mode and that of the burst in the long propagating mirror trajectory given by Eq.~(\ref{eq:mirtra}). 
 Since such an overlap is necessary for Wald's argument, our result implies that the explanation of the burst emission as a result of the monogamy of entanglement is incorrect.

In the moving mirror model, the mirror's velocity increases as long as the acceleration continues. Therefore, numerous particles must be emitted from the mirror in order to bring it to a complete stop. 
However, this burst emission is associated with the deceleration to stop the moving mirror, which does not mimic any phenomenon of the evaporation of a real black hole.
Since we have shown that the partner mode of Hawking radiation is not responsible for this  burst, the vacuum fluctuation scenario remains a promising framework.
%\textcolor{red}{ However, this means that  the moving mirror model does not mimic black hole evaporation because of emission of large amount of energy over the Planck  scale and violates assumption of the semi-classical treatment of gravity.
%Since the emission of the burst is not related to the vacuum fluctuation scenario,}  the vacuum fluctuation scenario remains a promising framework.

Although we did not address in this paper, we are interested in the following two issues; the first is the single-mode approximation adopted to obtain Eq.~(\ref{eq:singlemodeapprox}). In general, the partner is characterized by the general partner formula \cite{trevison2019spatially,nambu2023entanglement}, which is the relation between different modes of wave number and is nonlinear. This approximation becomes invalid when the weighting function $F(\omega)$ is not well localized. In such a case, there is an overlap between the profile of Hawking mode and  the profile of the partner mode, and we would not be able to discard Wald's argument based solely on the behavior of the profile of involved modes.
The second is the behavior of the renormalized entropy \cite{holzhey1994geometric} in the vacuum fluctuation scenario. For the long-propagating mirror trajectory followed by a uniform motion, the partner particle is returned as the vacuum fluctuation without energy gain from the mirror. However, following the Bianchi-Smerlak formula for entanglement entropy \cite{bianchi2014entanglement}, 
\begin{align}
	S(U):=-\frac{1}{12}\ln p'(U),
\end{align} 
wee see that the entanglement entropy does not decrease during the period of the inertial motion of the mirror.
Some researchers have also suggested improved definitions to recover unitarity by adding a constant to the renormalized entropy \cite{good2015signatures}. However, it is difficult to achieve the recovery of unitarity using these formulas. This is because the renormalization of the entropy involves the subtraction of the entanglement entropy associated with the static mirror trajectory, but this operation may also eliminate the contribution of vacuum fluctuations. Other approaches to renormalizing the entanglement entropy may be required to tackle this problem.
%These problems are left for our future work.

Last but not least, it would be interesting to make contact of our notions with experimentation. In particular, the international AnaBHEL (Analog Black Hole Evaporation via Lasers) Collaboration \cite{AnaBHEL2022} based on the Chen-Mourou proposal \cite{Chen2017Plasma} intends to generate a relativistic flying mirror through laser-plasma interaction. In this pursuit, the mirror's trajectory is determined by the variation of the background plasma density. A mirror trajectory that enables the emission of analog Hawking radiation can be designed via a proper density profile \cite{Chen2020Trajectory}. However, due to the complex nature of laser-plasma interaction, only a flying mirror with low and nontrivial reflectivity, in the sense of time and frequency dependence, can be feasibly generated, which deviates from the standard setup of a perfectly reflecting moving point mirror in (1+1) dimensions. For a given trajectory, such a low reflectivity setup results in the deviation of the particle number spectrum in comparison with that of a perfectly reflecting mirror \cite{Nistor2009Semi,Lin2021Particle,Lin2022Reflections,Lin2024Particle}. In addition, the energy flux emitted by the low-reflectivity moving mirror also deviates from that of the standard Fulling-Davies formula for two reasons: One is due to the nontrivial reflectivity effect for a given mode, and the other is due to the mixing of the reflected mode and the transmitted mode. Nevertheless, the total system can still be described in terms of a pure two-mode squeezed vacuum state \cite{Lin2024Particle}. Therefore, the formal construction of the detector/partner modes and their corresponding spatial profiles introduced in this paper should still be applicable to the setup in the AnaBHEL experiment.

While more detailed investigations are left for future work, it may be intuitively fair at this point to expect that, in the case of a suddenly stopping low-reflectivity mirror, the burst wave should exist. However, since the mirror has a low reflectivity, and since the partner mode of the Hawking mode is more energetic than the Hawking mode, it is highly possible that the partner mode will penetrate to the other side of the moving mirror. The AnaBHEL experiment may serve as a demonstration of the separate identity between the burst wave and the partner mode of the Hawking mode, since the burst and the partner are now on opposite sides of the moving mirror. Furthermore, since the partner mode is on the other side of the mirror, its flipped profile should also be on the other side of the burst wave, which again invalidates the entanglement monogamy viewpoint for the origin of the burst wave. To conclude, from either the viewpoints of the partner formula or the scenario of a low-reflective mirror, the final burst of the moving mirror seems to be \textit{disentangled} from both the Hawking mode's partner mode and the partner mode's flipped mode, therefore saving the vacuum fluctuation purification scenario from the extreme energy cost, i.e., \textit{quantum purity at a small price} is secured.

\begin{acknowledgements}
%  We would like to thank ?? for providing valuable insight on the subject. 
Y.N. was supported by JSPS KAKENHI (Grant No.~JP22H05257) and MEXT KAKENHI Grant-in-Aid for Transformative Research Areas A ``Extreme Universe"(Grant No.~24H00956).
Y.O. would like to take this opportunity to thank the “Nagoya University Interdisciplinary Frontier Fellowship” supported by Nagoya University and JST, the establishment of university fellowships towards the creation of science technology innovation, Grant Number JPMJFS2120.  MH
is supported by Grant-in-Aid for Scientific Research (Grant No.
21H05188, 21H05182, and JP19K03838) from the Ministry of
Education, Culture, Sports, Science, and Technology (MEXT),
Japan. 
PC was supported by Taiwan's National Science and Technology Council (NSTC) under project number 110-2112-M-002-031, and by the Leung Center for Cosmology and Particle Astrophysics (LeCosPA), National Taiwan University. KL appreciates the Elite Doctoral Scholarship jointly provided by the National Taiwan University (NTU) and Taiwan's National Science and Technology Council (NSTC).
\end{acknowledgements}

%%%%%%%%%%%%%%%%%%%%%%%%%%%%%%%%%%%%%%%%%%%%%%%%%%%%%%%
%%%%%%%%%%%%%%%%%%%%%%%%%%%%%%%%%%%%%%%%%%%%%%%%%%%%%%%
\appendix
%%%%%%%%%%%%%%%%%%%%%%%%%%%%%%%%%%%%%%%%%%%%%%%%%%%%%%%%
\section{Purification Partner of the Quantum Field}\label{sec:revpartner}
\begin{comment}\anno{Suggenstion on this part[by YN]:
\begin{itemize}
\item Delete general discussion on the partner formula.
\item Instead, use Rindler and Milne mode to explain the concept of the ``partner".
\item Use (37)(38) to show that the total state is a two-mode squeezed state under the single-mode approximation.
\end{itemize}}
\end{comment}
In the standard approach to quantum field theory in Minkowski spacetime \cite{weinberg1995quantum}, we typically expand the field operator in terms of plane wave modes. The choice of the plane-wave mode function is motivated by maintaining Lorentz invariance in the quantum field.
However, when acceleration is invoked, a plane-wave mode with respect to an inertial observer is no longer a plane-wave mode with respect to an accelerated observer since there is no Lorentz transformation relating these two frames.
In this case, the mode functions with respect to an accelerating observer $\{\psi_\Omega\}$ and those with respect to an inertial observer $\{\varphi_\omega\}$ are related through the Bogoliubov transformation:
\begin{align}
	\psi_\Omega=\int d\omega \left\{\alpha_{\Omega\omega}^*\,\varphi_\omega-\beta_{\Omega\omega}\,\varphi_\omega^*\right\}.
\end{align}
The coefficients $\alpha_{\Omega\omega}$ and $\beta_{\Omega\omega}$ are the Bogoliubov coefficients satisfying the following relation for each $\Omega$:
\begin{align}
	\int d\omega\,\{\alpha_{\Omega\omega}^*\alpha_{\Omega'\omega}-\beta_{\Omega\omega}\beta_{\Omega'\omega}^*\}=\delta_{\Omega\Omega'}.
\end{align}
Recall the expansion of the field operator
%%%
\begin{align}
	\hat{\phi}(x)=\int d\Omega\left\{\hat{A}_\Omega\,\psi_\Omega(x)+\hat{A}_\Omega{}^{\!\!\!\!\dag}\,\,\psi_\Omega{}^{\!\!*}(x)\right\}=\int d\omega\left\{\hat{a}_\omega\,\varphi_\omega(x)+\hat{a}_\omega{}^{\!\!\dag}\varphi_\omega{}^{\!\!\!*}(x)\right\},
\end{align}
where $\hat{A}_\Omega$ and $\hat{a}_\omega$ are annihilation operator with respect to $\{\psi_\Omega\}$ and $\{\varphi_\omega$\}, respectively,
we can rewrite the Bogoliubov transformation by using creation and annihilation operators:
\begin{align}
	\hat A_\Omega=\int d\omega \left\{\alpha_{\Omega\omega}\,\hat{a}_\omega+\beta_{\Omega\omega}{}^*\,\hat{a}_\omega{}^\dag\right\}.
\end{align}
Because the Bogoliubov transformation forms a group, we can also consider transformations between different accelerating observers.
Let us consider the case where the Bogoliubov transformation mixes only two different modes for simplicity; that is, we assume that $\alpha_{\Omega\omega}$ and $\beta_{\Omega\omega}$ have narrow peaks about $\omega=\omega_1$ and $\omega=\omega_2$, respectively. Then
\begin{align}
	\hat{A}:=\alpha\,\hat{a}_1+\beta^*\,\hat{a}_2{}^\dag.
\end{align}
where $\alpha$, $\beta$ satisfy $|\alpha|^2-|\beta|^2=1$. 
Since $\hat{a}_1$ and $\hat{a}_2$ are independent $[\hat{a}_1,\hat{a}_2]=[\hat{a}_1,\hat{a}_2{}^\dag]=0$,
the following commutation relation holds:
 \begin{align}
 	[\hat{A},\hat{A}^\dag]=1.
 \end{align}
General Bogoliubov transformations can be transformed into this form using mode transformations that do not alter the vacuum state and by redefining the local mode \cite{hotta2015partner}. The following discussion does not lose generality under this assumption. Since the Bogoliubov transformation does not change the number of modes, we have another counterpart to the Bogoliubov transformation:
\begin{align}
    \label{eq:partnerformula}
	\hat{B}:=\alpha\, \hat{a}_2+\beta^*\,\hat{a}_1^\dag,
\end{align}
and this operator satisfies the commutation relation
\begin{align}
	[\hat{B},\hat{B}^\dag]=1.
\end{align}
From their construction, two operators $\hat{A}$ and $\hat{B}$ are independent.
\begin{align}
	[\hat{A},\hat{B}]=[\hat{A},\hat{B}^\dag]=0,
\end{align}
and these operators $\hat{A}$, $\hat{B}$ annihilate different particle modes.
Since there are two pairs of the annihilation operators, we have two different vacuum states $|00\rangle_{12}$  and $|00\rangle_{AB}$ satisfying
\begin{align}
	0&=\hat{a}_1|00\rangle_{12}=\hat{a}_{2}|00\rangle_{12},\\
	0&=\hat{A}|00\rangle_{AB}=\hat{B}|00\rangle_{AB}.
\end{align}
These vacuum states are related as
\begin{align}
	|00\rangle_{12}=\sum_{n=0}^\infty\frac{(\tanh r)^n e^{in\varphi }}{\cosh r}|nn\rangle_{AB}
\end{align}
where the parameters $r,\varphi$ are introduced by $\tanh r=|\beta/\alpha|, \varphi=\arg(\beta^*/\alpha)$.
Tracing out $B$ degrees of freedom from the vacuum state $|00\rangle_{12}$, we have the mixed density matrix 
\begin{align}
	\hat{\rho}_A=\sum_{n=0}^\infty\frac{(\tanh^2 r)^{n} }{\cosh^2 r}|n\rangle\langle n|_A.
\end{align}
On the contrary, this mixed state for $A$ can be purified by $B$ degrees of freedom to produce the pure vacuum state $|00\rangle_{12}$. In this sense, the mode $\hat{B}$ that appears as the counterpart of the Bogoliubov transformation is called the partner mode of the mode $\hat{A}$, and Eq.~(\ref{eq:partnerformula}) is called the partner formula.
%%%%%%%%%%%%%%%%%%%%%%%%%%%%%%%%%%%%%%%%%%%%%%%%%%%%%%%%%%%%%%%%%
%%%%%%%%%%%%%%%%%%%%%%%%%%%%%%%%%%%%%%%%%%%%%%%%%%%%%%%%%%%%%%%%%
\section{Nonlinearlity of the Partner Formula and the Single Mode Approximation}\label{sec:ap-B}
For a mode defined by independent annihilation operators $\hat a_1, \hat a_2$,
\begin{align}
	\hat{A}_1:=\alpha_1\,\hat{a}_1+\beta^*_1\,\hat{a}_2^\dag,\quad|\alpha_1|^2-|\beta_1|^2=1,
\end{align}
its partner is given by
%%%
\begin{align}
	\hat{A}_{1P}:=\alpha_1\,\hat{a}_2+\beta_1^*\,\hat{a}_1^\dag.
\end{align}
%%%
A set of annihilation operators $(\hat A_1, \hat A_{1P})$ constitutes a two mode pure state.
Now let us consider another set of modes defined by the following Bogoliubov transformation using independent annihilation operators $\hat b_1, \hat b_2$:
%%%
\begin{align}
	\hat{A}_2:=\alpha_2\,\hat{b}_1+\beta^*_2\,\hat{b}_2^\dag,\quad|\alpha_2|^2-|\beta_2|^2=1.
\end{align}
%%%
Then the partner mode for this mode is given by
\begin{align}
	\hat{A}_{2P}:=\alpha_2\,\hat{b}_2+\beta_2^*\,\hat{b}_1^\dag.
\end{align}
%%%
We assume that $(\hat{a}_1$,  $\hat{a}_2)$ and $(\hat{b}_1$, $\hat{b}_2$) are independent of each other.
From two annihilation operators $\hat{A}_1$ and $\hat{A}_2$, we can define a new annihilation operator by
%%%
\begin{align}
	\hat{A}=\cos\theta\hat{A}_1+\sin\theta\hat{A}_2.
\end{align}
%%%
The annihilation operator $\hat{A}$ can be rewritten as
\begin{align}
	\hat{A}\equiv\alpha\,\hat{a}_{\parallel}+\beta^*\,\hat{a}_\perp^\dag
\end{align}
%%%
where
%%%
\begin{align}
	\alpha&:=\sqrt{\cos^2\theta|\alpha_1|^2+\sin^2\theta|\alpha_2|^2}
	, \quad\beta:=\sqrt{\cos^2\theta|\beta_1|^2+\sin^2\theta|\beta_2|^2},\\
	\hat{a}_\parallel&:=\frac{1}{\alpha}\left((\cos\theta)\alpha_1\hat{a}_1+(\sin\theta)\alpha_2\hat{b}_1\right),
	\quad\hat{a}_\perp:=\frac{1}{\beta}\left((\cos\theta)\beta_1\hat{a}_2+(\sin\theta)\beta_2\hat{b}_2\right),
\end{align}
%%%
and its partner is given by
\begin{align}
	\hat{A}_P:=\alpha\,\hat{a}_\perp+\beta^*\,\hat{a}_\parallel.
\end{align}
%%%
Since the square of $\cos\theta$, and $\sin\theta$ appear in the formula of $\hat{A}_P$, the relationship between these partners is generally nonlinear:
%%%
\begin{align}
	\hat{A}_P\ne\cos\theta\hat{A}_{1P}+\sin\theta\hat{A}_{2P}.
\end{align}
%%%
However, if the two modes $\hat{A}_1$ and  $\hat{A}_2$ share almost the same Bogoliubov coefficients (\textit{i.e.} $\beta_1\simeq\beta_2$), that is applicable for the single mode approximation, we obtain the following linear relationship between partners:
%%%
\begin{align}
	\hat{A}_P=\cos\theta\hat{A}_{1P}+\sin\theta\hat{A}_{2P}.
\end{align}
%%%
For a mode defined by the linear combination of more than three independent annihilation operators,
\begin{align}
    \hat{A}:=\sum_i f_i\hat{A}_i,\hspace{0.5cm} \sum_i |f_i|^2=1,
\end{align}
%%%
by repeating the above procedure within the single mode approximation,
we obtain the partner mode, 
\begin{align}
    \hat{A}_P:=\sum_i f_i\hat{A}_{iP}.
\end{align}
For the case of Rindler mode, the Bogoliubov coefficients are given as functions of frequency $\omega$, and if the weighting function $F(\omega)$ has a sharp peak at some frequency,  the relationship between partners becomes  a linear one.
Recalling that the partner mode of the Rindler mode $\hat{a}^{\text{I}}_\omega$ is the Milne mode $\hat{a}^{\text{II}}_\omega$, the partner mode of the Eq.~(\ref{eq:annidet}) can be approximated as
\begin{align}
    \hat{a}_P=\int\dd\omega f(\omega)\hat{a}^{\text{II}}_\omega.
\end{align}
This is the meaning of the single mode approximation adopted in Eq.~(\ref{eq:singlemodeapprox}). 

%%%%%%%%%%%%%%%%%%%%%%%%%%%%%%%%%%%%%%%%%%%%%%%%%%%%%%%%
\section{Energy of the Partner Mode}\label{ap-a}
Let us consider the energy of the partner particles. We can estimate the energy of the partner particles by  
%%%
\begin{align}
	E_P:=\left\langle\hat{a}_P^\dag\hat{a}_P+\frac{1}{2}\right\rangle_{in}.
\end{align}
%%%
Rewriting this formula using canonical modes, 
\begin{align}
	E_P&=\left\langle\frac{\hat{Q}_P^2+\hat{P}_P^2}{2}\right\rangle+\frac{1}{2}\nn
	&=\frac{1}{2}\int dU\dd {U'} \left(q_P(U)q_P(U')+p_P(U)p_P(U')\right)\langle\hat{\Pi}^R(U)\hat{\Pi}^R(U')\rangle_{in}+\frac{1}{2}\nn
	&=\frac{1}{2}\int dU \dd {U'} \frac{p'(U)p'(U')(q_P(U)q_P(U')+p_P(U)p_P(U'))}{(p(U)-p(U')-i\varepsilon)^2}+\frac{1}{2}\nn
	&=\frac{1}{2}\int dU \dd {U'} \left\{\left(\frac{p'(U)p'(U')}{(p(U)-p(U')-i\varepsilon)^2}-\frac{1}{(U-U'-i\varepsilon)^2}\right)+\frac{1}{(U-U'-i\varepsilon)^2}\right\}(q_P(U)q_P(U')+p_P(U)p_P(U'))+\frac{1}{2}.\notag
\end{align}
%%%
The first term in the curly bracket equals to the Fulling-Davies flux formula and this corresponds to the renormalized energy flux $\langle\hat{T}_{UU}\rangle_{ren}$ radiated from moving mirror, and the second term corresponds to the contribution from vacuum fluctuation of the quantum field.
The energy flux radiated from the mirror is zero during the inertial motion of the mirror, and we can choose detector mode so that the profile of the partner mode is approximately compactified in the region where the energy flux of the mirror vanishes for suddenly stopping and long-propagating mirror trajectories.
Therefore, the energy of the partner mode is nothing more than the energy of the vacuum fluctuation, as we have expected.
%%%%%%%%%%%%%%%%%%%%%%%%%%%%%%%%%%%%%%%%%%%%%%%%%%%%%%%
%%%%%%%%%%%%%%%%%%%%%%%%%%%%%%%%%%%%%%%%%%%%%%%%%%%%%%%%

%%%%%%%%%%%%%%%%%%%%%%%%%%%%%%%%%%%%%%%%%%%%%%%%
\bibliography{mirror} %hoge.bibから拡張子を外した名前
%\bibliographystyle{junsrt} %参考文献出力スタイル
%%%%%%%%%%%%%%%%%%%%%%%%%%%%%%%%%%%%%%%%%%%%%%%

\end{document}